\documentclass[showpacs,aps,prl,superscriptaddress]{revtex4}

\usepackage{graphicx}
\usepackage{dcolumn}
\usepackage{amsmath}
\usepackage{epsfig}

\newcommand{\TeV}{\ensuremath{\mathrm{Te\kern -0.1em V}}}
\newcommand{\TeVc}{\ensuremath{\mathrm{Te\kern -0.1em V\!/}c}}
\newcommand{\TeVcc}{\ensuremath{\mathrm{Te\kern -0.1em V\!/}c^2}}
\newcommand{\GeV}{\ensuremath{\mathrm{Ge\kern -0.1em V}}}
\newcommand{\GeVc}{\ensuremath{\mathrm{Ge\kern -0.1em V\!/}c}}
\newcommand{\GeVcc}{\ensuremath{\mathrm{Ge\kern -0.1em V\!/}c^2}}
\newcommand{\MeV}{\ensuremath{\mathrm{Me\kern -0.1em V}}}
\newcommand{\MeVc}{\ensuremath{\mathrm{Me\kern -0.1em V\!/}c}}
\newcommand{\MeVcc}{\ensuremath{\mathrm{Me\kern -0.1em V\!/}c^2}}

\def\figurebox#1#2#3{%
    \def\arg{#3}%
    \ifx\arg\empty
    {\hfill\vbox{\hsize#2\hrule\hbox to #2{\vrule\hfill\vbox to #1{\hsize#2\vfill}\vrule}\hrule}\hfill}%
    \else
    {\hfill\epsfbox{#3}\hfill}%
    \fi}

\begin{document}

\begin{flushleft}
\end{flushleft}

\title
{
{\large \bf 
Things that go bump in the night: \\
From   $J/\psi\phi$ to  other  mass spectrum 
} 	
}
	
\author{
Kai Yi
}

\address{
Department of Physics and Astronomy, University of Iowa,\\
Iowa City, Iowa 52242-1479, USA\\
email: yik@fnal.gov
}

\date{\today}

\begin{abstract}

This article summarizes a brief development of exotic mesons with Vector-Vector (VV) final states 
starting  from the $J/\psi\phi$ mass spectrum, as well as  
extensions to  other  final states for  strong-interaction physics and 
physics beyond standard model. These VV final states are very suitable 
to be studied at CEPC through associate production and two-photon production, 
as well as a visionary low energy photon collider.

\end{abstract}

\pacs{
 }     

\maketitle

\section{Introduction}

The existence of potential multiple structures in the $J/\psi\phi$ mass spectrum~\cite{cdfjpsiphi} 
was in question several years ago, reference~\cite{kyreview} summarized the status of $J/\psi\phi$ 
mass spectrum and the Vector-Vector  (VV) final states by then, 
and proposed to search for structures near threshold in the VV systems composed entirely of c 
quarks and b quarks. 
Now the first two structures in the $J/\psi\phi$ mass spectrum, renamed as $X(4140)$ and $X(4274)$ 
by PDG, have been confirmed or 
re-confirmed by CMS~\cite{cmsjpsiphi}, D0~\cite{d0jpsiphi} and LHCb~\cite{lhcbjpsiphi}, in addition   
the LHCb collaboration announced two more new structures, named as $X(4500)$ and $X(4700)$ in the 
same mass spectrum. 
Both BaBar and BES collaborations searched the  $J/\psi\phi$ mass spectrum without 
significant signals~\cite{babarbesjpsiphi}. These states are the first of such kind of exotic meson 
candidates that have no light quark components, and added an important piece to the 
exotic meson family.  There are total five structures reported in the $J/\psi\phi$ mass spectrum 
adding the $X(4350)$ spotted in the two-photon process at Belle experiment~\cite{bellex4350}. 
The $J^{PC}$ possibilities of $1^{-+}$, $1^{++}$ and $3^{-+}$ are excluded for  $X(4350)$  
inferred from its production process. Assuming $X(4350)$ is real, it is suppressed 
in the B decays  in the case of a spin 2 state.  There is also large difference for the 
width of the $X(4140)$ measued by LHCb experiment and other experiments~\cite{PDG}.
More statistics from LHC and Belle II will help 
to make things clear.  These states demand  a through theoretical understanding eventually.  
Possible interpretations have been  developed over years, however, we do not have a perfect model 
or explanation for these $J/\psi\phi$ structures.

A natural extension for the $J/\psi\phi$ mass spectrum would be the $\psi(2S)\phi$ system. 
The CMS collaboration has observed 
the relevant decays--$B^{\pm}\rightarrow \psi(2S)\phi K^{\pm}$, however, 
due to limited statistics and the limited available phase space allowed for $\psi(2S)\phi$ mass 
in the B decays, CMS did not exam the $\psi(2S)\phi$ mass spectrum in this decay~\cite{cmspsi2sphi}.
Other VV systems, include $\psi \psi$, $\psi\Upsilon$, $\Upsilon\Upsilon$,  
can be further explored at LHC with large data sample, where $\psi$ and $\Upsilon$ 
can be either  on-shell or off-shell.   
The leptonic decays of these vector states provide many advantages for investigation, 
especially at ATLAS and CMS experiments that have no good hadron particle identification 
power--but both experiments have excellent lepton identification performance and momentum resolution, 
as well as excellent mass resolution. 
The production cross section of  $J/\psi J/\psi$ has been measured at LHC experiments 
at various energies~\cite{jpsijpsicxatlhc}, and recently CMS measured the production 
cross section of $\Upsilon\Upsilon$ at $\sqrt{S}=8$ TeV for the first time~\cite{cmsupsiupsicx}. 
The production of $J/\psi \Upsilon$ at $\sqrt{S}=8$ TeV has been preliminarily investigated 
in a thesis work~\cite{kamuranthesis}. These benchmark measurements show evidence 
that there are enough statistics to search for possible structures in these 
VV systems composed entirely of c quark and b quarks.

A recent preliminary investigation through a thesis work~\cite{suleyman} shows an evidence of a 
structure at 18 GeV in the $\Upsilon(1S)\ell^+\ell^-$ final states with a width   
less than 150 MeV. The observed mass is several hundreds MeV below 
the threshold of double $\eta_b$ 
mass, which is the first one among the observed X/Y/Z states, the others are either above or near  
the threshold of two particles. Assuming it as a real physics effect, 
one possibility is that this is a new state decaying to  
one on-shell and one off-shell $\Upsilon(1S)$, which is another VV system with V being  
on-shell or off-shell. This article focuses on possibilities of investigating the  
VV system at various existing or visionary colliders in the world.

\section{Motivation}

A recent summary of the discovered X/Y/Z states in the last decade 
can be found at Ref.~\cite{PDG}, and the recent theoretical developments of understanding 
of these X/Y/Z states can be  found at Ref.~\cite{rosnerreview}.  
The nature of these states is far away from a complete understanding, despite of the fact 
that many models  such as hadrocharmonium, molecule, CUSP etc have been proposed to explain.  
As mentioned in the introduction section, 
there is one kind of candidates that decays to VV mesons without light quark 
component among the newly discovered exotic candidate zoo, where V can be either an on-shell 
or off-shell vector meson. These VV candidates are not understood and there is little 
information known from both production and decay aspects. On the other hand, 
these VV system can be produced through two-photon collision, and this article provides 
preliminary investigation of decaying kinematic  distributions  
at CEPC~\cite{cepc} and a visionary photon collider. 
In addition, possible associate production at CEPC is also discussed.


As for the 18 GeV excess reported in the thesis work~\cite{suleyman}, 
several possibilities exist. Assuming this is due to real physics effects, a simple naive  
explanation would be a bound state composed of tetra-bottom quark which is analog to 
positronium molecule state~\cite{positroniummolecule}.  Theoretical development along this direction 
can be found at Ref~\cite{4b}. However, a recent lattice QCD calculation denies the existence 
of a tetra-bottom quark state~\cite{4blattice}.
Other possibilities include a light scalar or pseudo-scalar state.
Nevertheless, the  region below the Z mass has never been fully explored, it is worth 
to investigate as a general case with some efforts. 
This article discusses possible exploration in this mass region at various 
existing or visionary colliders, the focus is on the kinematic distributions 
as a general phase space decay.

\subsection{Possible  decay channels to explore}

As stated above, it is not thoroughly investigated for the  region 
below the Z mass, there is potential for new composite or elementary particles existing 
in this region, thus we assume there are  possible new states in this region,  
and study the kinematics in various decay channels, and in various facilities. 
Assuming a  new state existing below the Z mass, the possible decay channels are  
different and the relative ratios among various channels are also different, depending 
on what it is.  For instance, a typical hadronic state such as a tetra-bottom (tetra-charm) quark 
state would prefer to decay into bottomonium (charmonium) related final states, on the other hand,  
a typical new state that is an extension of standard model may choose to decay 
to $\gamma\gamma$ or $jj$. There are possibilities that both hadronic states and elementary states 
can exist in this mass region, thus we consider the following decay channels as a general case:
$VV$ where V can be an on-shell or off-shell vector meson; $V \gamma$;
$\ell^+\ell^-$; $\gamma\gamma$; $jj$  etc.  The possible new 
state masses assumed in this article are 18.5, 30 and 50 GeV.

\begin{itemize}

\item V+V, where V is an on-shell or off-shell vector meson with $J^{PC}=1^{--}$ such as 
$J/\psi$, $\psi(2S)$, $\Upsilon(nS)$ etc. These channels can be done at the current ATLAS/CMS experiments  
with V decays to dilepton final states. Depends on the lepton kinematics at different mass 
regions, this can be difficult for low momentum leptons due to pileup  
effects.   For instance, the peak luminosity at CMS varies from 1 Hz/nb in 2011 
to 22 Hz/nb in 2017~\cite{cmslumi}, and it is going to be much higher in the future which makes the 
pileup effect more serious. It is possible to improve the performance with improved 
detector technology such as  timing separation for the charged particles. 
On the other hand, these channels can be fully investigated at 
LHCb when enough data samples are accumulated by taking advantage of its relative  
low peak luminosity (lower pileup).

\item V+$\gamma$, where $\gamma$ can be a reconstructed photon object, or a converted photon 
reconstructed using trackers, or a virtual gamma decay. Considering the background rate, 
fake rate and low efficiency for low momentum photons, converted photons seem more plausible 
for ATLAS/CMS by taking advantage of the vertex information at the conversion point.

\item $\gamma\gamma$, $q\bar{q}$, $\ell^+\ell^-$ where $\ell=\tau,\mu,e$. These channels 
are generally difficult to do at the LHC facilities due to trigger restriction, reconstructing 
challenge and high background rate.  However, they can be performed relatively easy at 
other facilities such as CEPC.  Several examples will be demonstrated 
in the following sections.

\end{itemize}

\clearpage
\newpage

\section{ Investigation at CEPC facility }

The Circular Electron Positron Collider (CEPC) is a collider being proposed to 
study the Higgs boson in precision at the collision energy of 250 GeV~\cite{cepc}.
However, this facility can be used to study many other physics, and this article 
discusses the possible investigation of the VV system 
at  this $e^+e^-$ collider.  The physics tool used here is Pythia8~\cite{pythia8}, 
and the detector simulation tool used is Delphes~\cite{delphes}, the specific 
configuration of the detector used is the embedded CEPC detector in Delphes tool. 
All decays are simply phase space kinematics.
The VV system can not be produced through $e^+e^-$  annihilation because 
$e^+e^-$ typically annihilate to a virtual photon which is a $1^{--}$
object.  The VV system, which may come from a new state decays, 
can be produced at CEPC through two processes: 
1) associate production--it can be produced associated with an object such as 
a Z or a photon.
2) two-photon process--both electron and positron radiate a virtual photon and the two photons collide 
to produce it.
In principle, this can also be searched and studied inclusively, but we skip 
the discussion.

There is no detailed cross section information available yet for various production 
mechanisms, though 
it can be inferred from Ref.~\cite{ilya} for two-photon production, 
the author is looking for possible collaboration with theorists to investigate 
production rate in a through way with different assumptions.

\subsection{Associate production}

There are two ways to search for a new object with associate production at CEPC. 
One is to look for the recoiled mass against a fully reconstructed object such as a Z boson, 
another way is to investigate a specifically reconstructed final state, 
which can include a fully reconstructed associate object or can be inclusive. 
This section discusses the two methods to investigate new objects. 


\subsubsection{With Recoil technique}

\begin{figure}[b]
\centerline{\includegraphics[width=15.0cm]{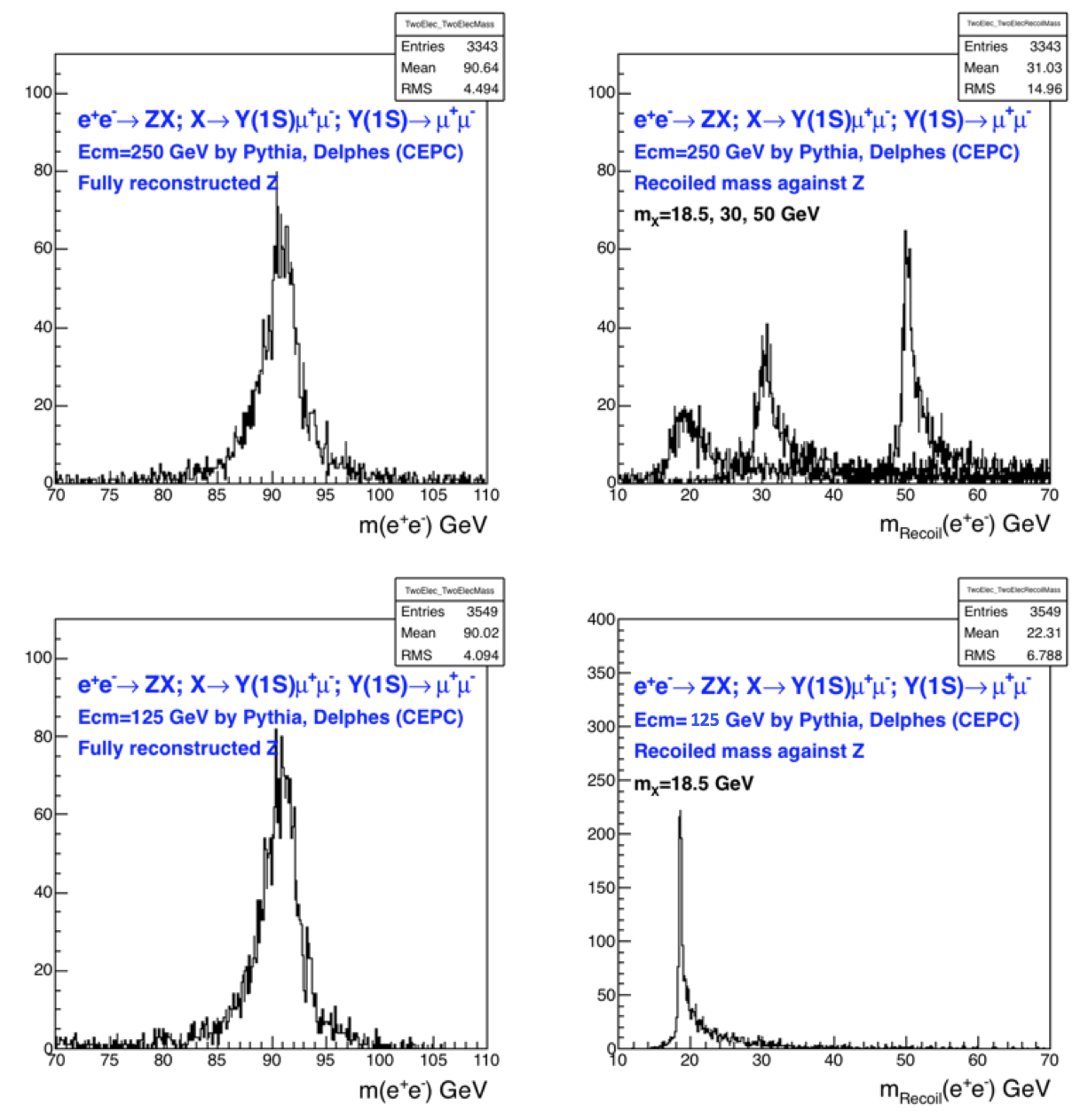}}
\caption{
The fully reconstructed Z mass and its recoiled mass against Z 
particle at CEPC with different collision energy by assuming 
X particle mass as 18.5, 30 and 50 GeV produced in the process 
of $e^+e^- \rightarrow Z X; X\rightarrow Y(1S) \mu^+\mu^-; Y(1S)\rightarrow \mu^+\mu^-$.
These events are simulated by Delphes with CEPC configuration.
\label{cepcfig1}}
\end{figure}

The benefit to investigate the recoiled mass is that it includes  
the object's all final states, which provides large rate to be more plausible  
to observe, but it is affected by mass resolution and background.  
The branching fraction for a specific channel can also be measured 
if a specific channel is reconstructed in the recoil side. 
Figure~\ref{cepcfig1} shows the fully reconstructed Z($e^+e-$) mass and 
the recoiled mass against the Z boson at different collision energies  
with different object masses: 18.5, 30 and 50 GeV. 
The Z mass resolution is more or less the same at different collision energies  
as shown in Fig.~\ref{cepcfig1} (top and bottom left).
Figure~\ref{cepcfig1} (top right) shows the recoiled 
mass distributions with different  mass values at collision energy of 250 GeV.  
The mass resolution varies from several GeV at low mass region  
to less than 1 GeV at higher mass region.
Note that the mass resolution becomes better with higher mass objects 
due to the squeezed phase space.
The mass resolution can be greatly improved in the low mass region 
with lower collision energy as shown in Fig.~\ref{cepcfig1} (bottom right), 
this can be done with the designed low collision energy at the beginning.

As we will discuss in the next section, the new object can also be fully reconstructed 
with a specific channel. Combining with the information with recoiled mass measurement, 
the branching fraction for a specific channel can be measured.

\subsubsection{With fully reconstructed objects}

Here we choose the following specific channels as examples 
to demonstrate  their kinematic distributions at CEPC with phase space decays.

\begin{figure}[b]
\centerline{\includegraphics[width=15.0cm]{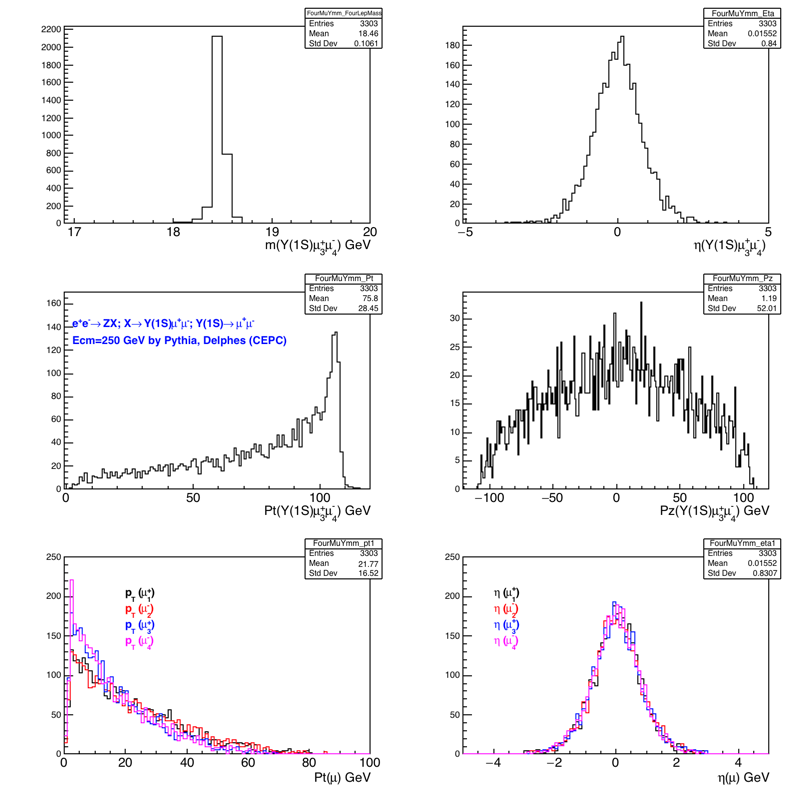}}
\caption{
The features of the four-muon system produced in 
$e^+e^- \rightarrow Z X; X\rightarrow Y(1S) \mu^+\mu^-; Y(1S)\rightarrow \mu^+\mu^-$ process  
at CEPC: 
(top left) The invariant mass of the four muons; 
(top right) The $\eta$ distribution of the four muons;
(middle left)$p_T$ of the four muons;
(middle right) $p_Z$ of the four muons;
(bottom left) $p_T$ of each muon.
(bottom right) $\eta$ of each muon.
These events are simulated by Delphes with CEPC configuration.
\label{cepcfig2}}
\end{figure}

\begin{figure}[b]
\centerline{\includegraphics[width=15.0cm]{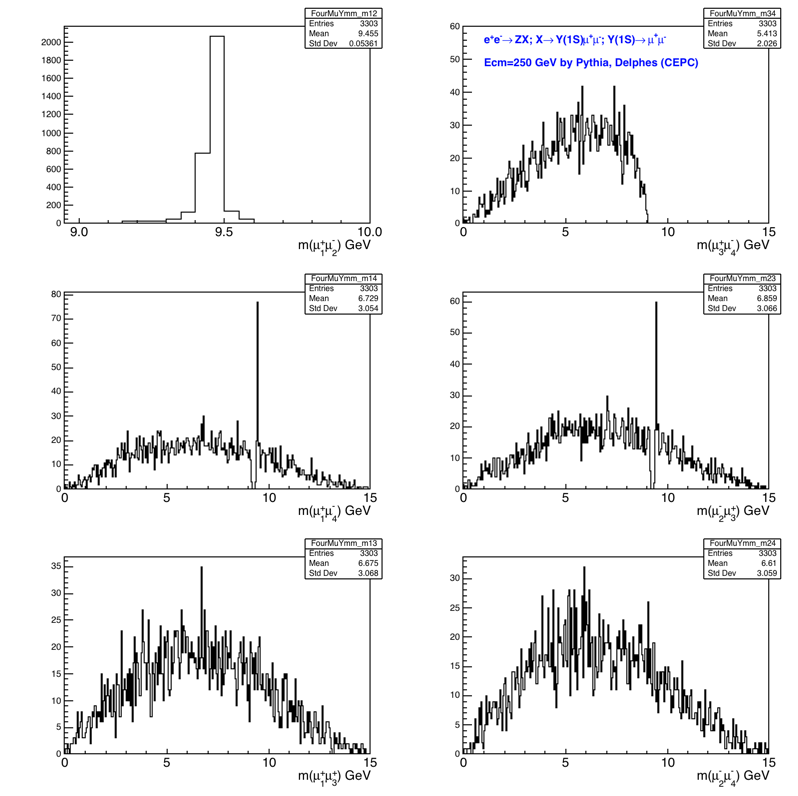}}
\caption{
The six dimuon mass distributions of the four-muon system in the $e^+e^- \rightarrow Z X; X\rightarrow Y(1S) \mu^+\mu^-; Y(1S)\rightarrow \mu^+\mu^-$  process 
at CEPC.
\label{cepcfig3}}
\end{figure}

\begin{figure}[b]
\centerline{\includegraphics[width=15.0cm]{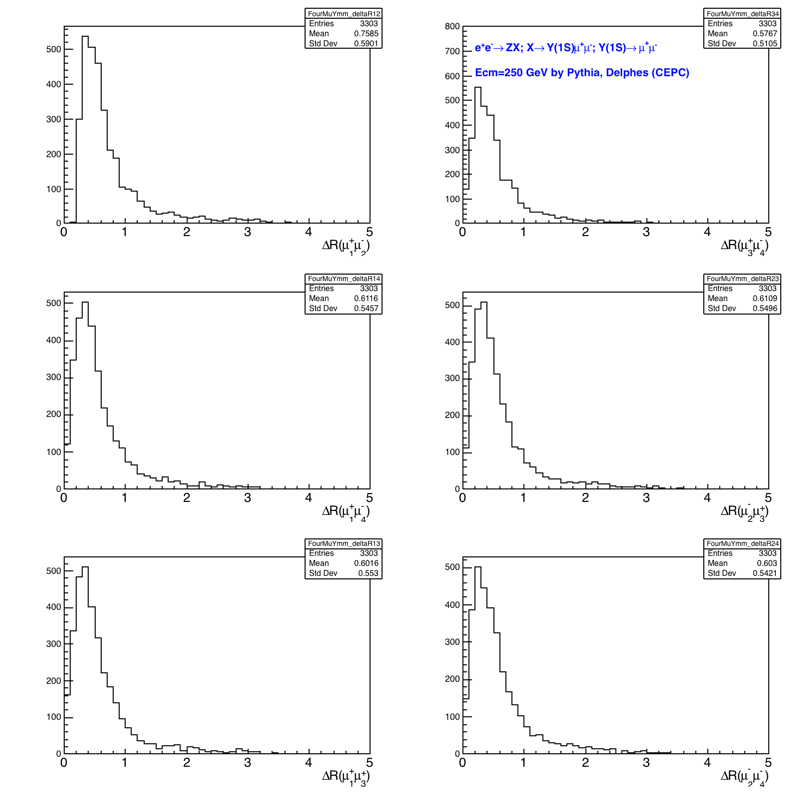}}
\caption{
The six $\Delta R$ distributions of the four-muon system in the $e^+e^- \rightarrow Z X; X\rightarrow Y(1S) \mu^+\mu^-; Y(1S)\rightarrow \mu^+\mu^-$  process 
at CEPC.
\label{cepcfig4}}
\end{figure}

\begin{figure}[b]
\centerline{\includegraphics[width=15.0cm]{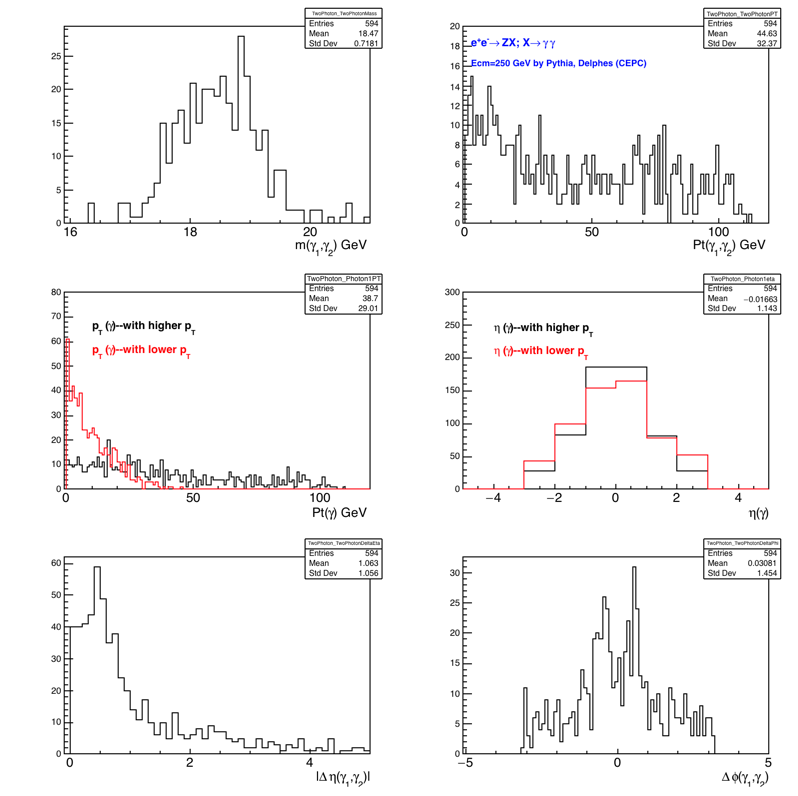}}
\caption{
The features of the $\gamma\gamma$ system in the $e^+e^- \rightarrow Z X; X\rightarrow \gamma\gamma$ process 
at CEPC: 
(top left) The di-photon mass distribution;
(top right) The $p_T$ distributions of the di-photon;
(middle left) The $p_T$ distribution of each photon;
(middle right) The $\eta$ distribution of each photon;
(bottom left) The $\Delta \eta$ distribution of the di-photon;
(bottom right) The $\Delta \phi$ distribution of the di-photon.
These events are simulated by Delphes with CEPC configuration.
\label{cepcfig5}}
\end{figure}

\begin{figure}[b]
\centerline{\includegraphics[width=15.0cm]{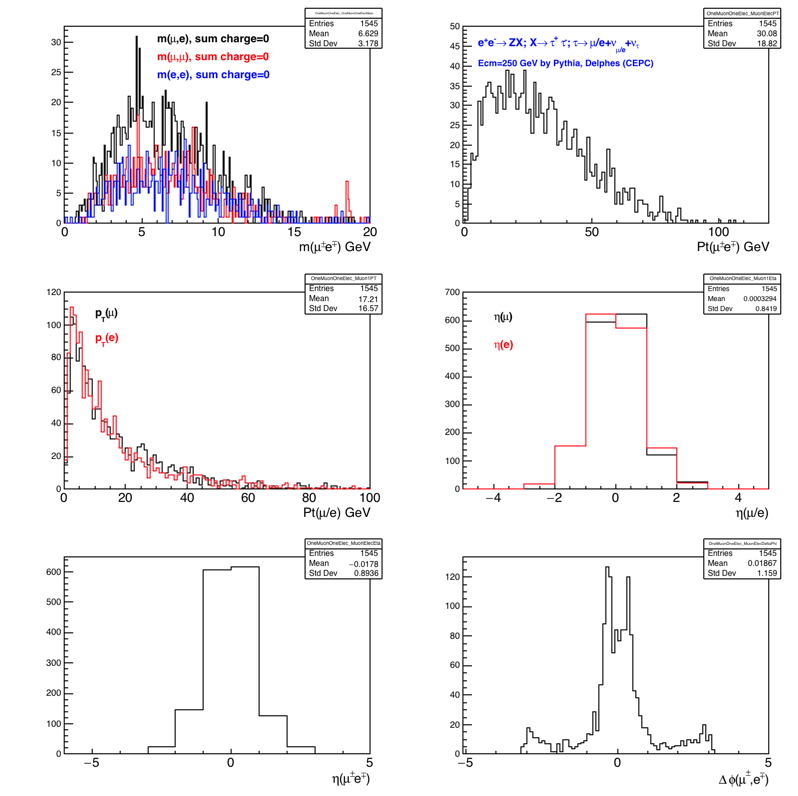}}
\caption{
The features of the $\mu^{\pm}e^{\mp}$ ($\tau^{\pm}\tau^{\mp}$) system 
in the $e^+e^- \rightarrow Z X; X\rightarrow \tau^{\pm}\tau^{\mp}; \tau^{\pm}\rightarrow\ell^{\pm}+anything-else, 
\tau^{\mp}\rightarrow \ell^{\mp}+anything-else$, where $\ell=\mu, e$ 
process at CEPC: 
(top left) The $\mu^{\pm}e^{\mp}$/$\mu^{\pm}\mu^{\mp}$/$e^{\pm}e^{\mp}$ mass distributions;
(top right) The $p_T$ distributions of $\mu^{\pm}e^{\mp}$;
(middle left) The $p_T$ distribution of muon and electron;
(middle right) The $\eta$ distribution of muon and electron;
(bottom left) The $\eta$ distribution of $\mu^{\pm}e^{\mp}$;
(bottom right) The $\Delta \phi$ distribution  between muon and electron.
These events are simulated by Delphes with CEPC configuration.
\label{cepcfig6}}
\end{figure}

\begin{itemize}

\item $e^+e^- \rightarrow Z X; X\rightarrow Y(1S) \mu^+\mu^-; Y(1S)\rightarrow \mu^+\mu^-$.
Figure~\ref{cepcfig2} shows the reconstructed mass, $\eta$, $p_T$, $p_Z$  distributions 
for a potential 18 GeV object and its decay particles for this process. 
Due to the recoiling boost from Z boson in the transverse plane, this object has large transverse 
momentum up to 110 GeV, this feature can be used to select events to reduce background 
in the final state. The muons in the final states are mostly distributed in the detector's central 
region. Figure~\ref{cepcfig3} and ~\ref{cepcfig4} show the 
six dimuon mass distributions and the six $\Delta R$ distributions 
between the six dimuon pairs.  The four muons are close to each other against the 
Z boson.  As one can see, there are mispaired real upsilon in the other dimuon 
pairs: (1,4) and (2,3), where $\mu_1$ and $\mu_2$ form the on-shell $\Upsilon(1S)$ particle 
and $\mu_3$ and $\mu_4$ form the second dimuon pair.

\item $e^+e^- \rightarrow Z X; X\rightarrow \gamma\gamma$.
Figure~\ref{cepcfig5} shows the reconstructed mass, $p_T$, $\eta$, $\Delta \eta$ and 
$\Delta \phi$  distributions for a potential 18 GeV object and its decaying particles. 
Due to the boost from Z boson in the transverse plane, this object has large transverse 
momentum up to 110 GeV, and the high $p_T$ photon has relative flat transverse 
momentum up to 100 GeV, which can be used to optimize  the signal 
in the final state. 
The photons in the final states are mostly distributed in the detector's central 
region with close distance in $R$ and $\phi$ spaces.

\item $e^+e^- \rightarrow Z X; X\rightarrow \tau^{\pm}\tau^{\mp}; \tau^{\pm}\rightarrow\ell^{\pm}+anything-else, \tau^{\mp}\rightarrow \ell^{\mp}+anything-else$, where $\ell=\mu, e$.
Figure~\ref{cepcfig6} shows the reconstructed $\mu^{\pm}e^{\mp}$, $\mu^{\pm}\mu^{\mp}$ 
and $e^{\pm}e^{\mp}$
 mass distributions, $p_T$, $\eta$, $\Delta \eta$ and 
$\Delta \phi$  distributions for the dilepton system. 
Due to the missing neutrino particles, the reconstructed dilepton masses are not peaked 
at 18.5 GeV but with a wide mass spectrum below 18.5 GeV. 
The transverse momentum is soften by missing neutrino particles in the final state.
The dilepton system  and each lepton are  mostly distributed in the detector's central 
region with close distance in $R$ and $\phi$ spaces.  This channel may be only possible in 
$e^+e^-$ machines due to its low background.

\end{itemize}

\clearpage
\newpage

\subsection{Two-photon production}

Another possible production for potential new states at CEPC is via the two-photon process:
$e^+e^- \rightarrow e^+e^- \gamma^* \gamma^*,  \gamma^* \gamma^* \rightarrow X$.
A striking feature of the two-photon process is that the reconstructed object 
shows ``zero'' transverse momentum and it is typically used as a selection criteria  
to separate  signal events from background events. However, 
particles with  $J^{PC}$  of $1^{-+}$, $1^{++}$ and $3^{-+}$ are restricted in two-photon 
process. 
Three specific channels at CEPC produced through two-photon process 
are discussed below.

\begin{figure}[b]
\centerline{\includegraphics[width=15.0cm]{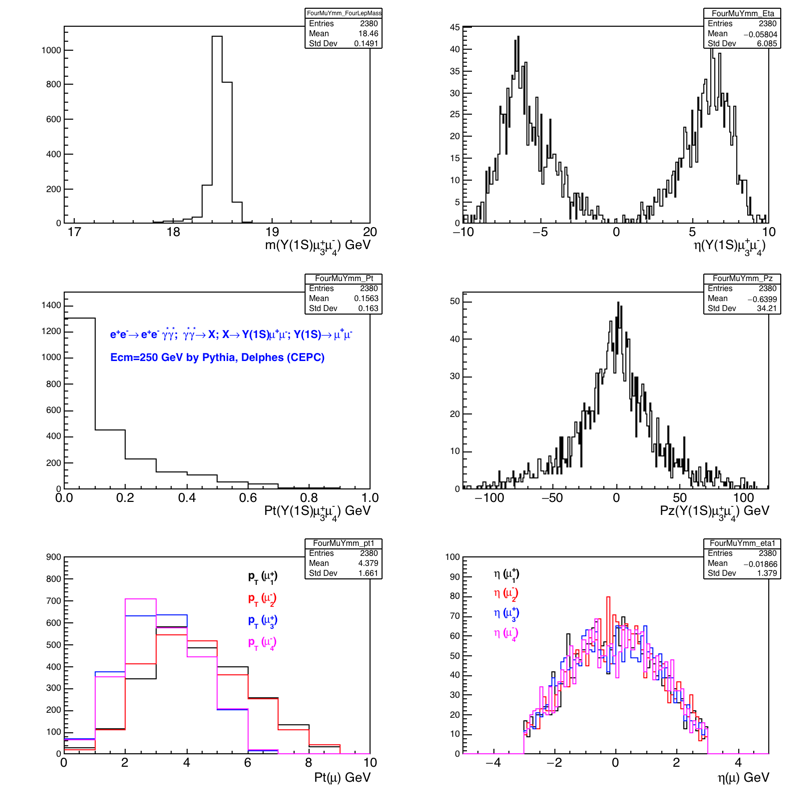}}
\caption{
The features of the four-muon system in the  $e^+e^- \rightarrow e^+e^- X; X\rightarrow Y(1S) \mu^+\mu^-; Y(1S)\rightarrow \mu^+\mu^-$ process 
at CEPC: 
(top left) The invariant mass of the four muons; 
(top right) The $\eta$ distribution of the four muons;
(middle left)$p_T$ of the four muons;
(middle right) $p_Z$ of the four muons;
(bottom left) $p_T$ of each muon.
(bottom right) $\eta$ of each muon.
These events are simulated by Delphes with CEPC configuration.
\label{cepcfigtvp1}}
\end{figure}

\begin{figure}[b]
\centerline{\includegraphics[width=15.0cm]{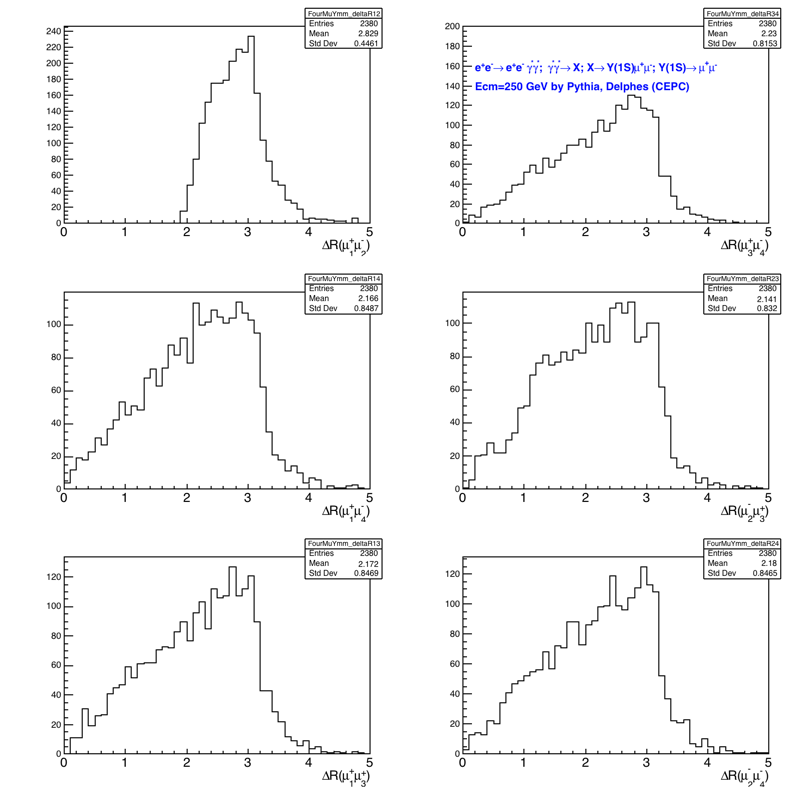}}
\caption{
The six $\Delta R$ distributions of the four-muon system in the  $e^+e^- \rightarrow e^+e^- X; X\rightarrow Y(1S) \mu^+\mu^-; Y(1S)\rightarrow \mu^+\mu^-$  
process 
at CEPC.
\label{cepcfigtvp2}}
\end{figure}

\begin{figure}[b]
\centerline{\includegraphics[width=15.0cm]{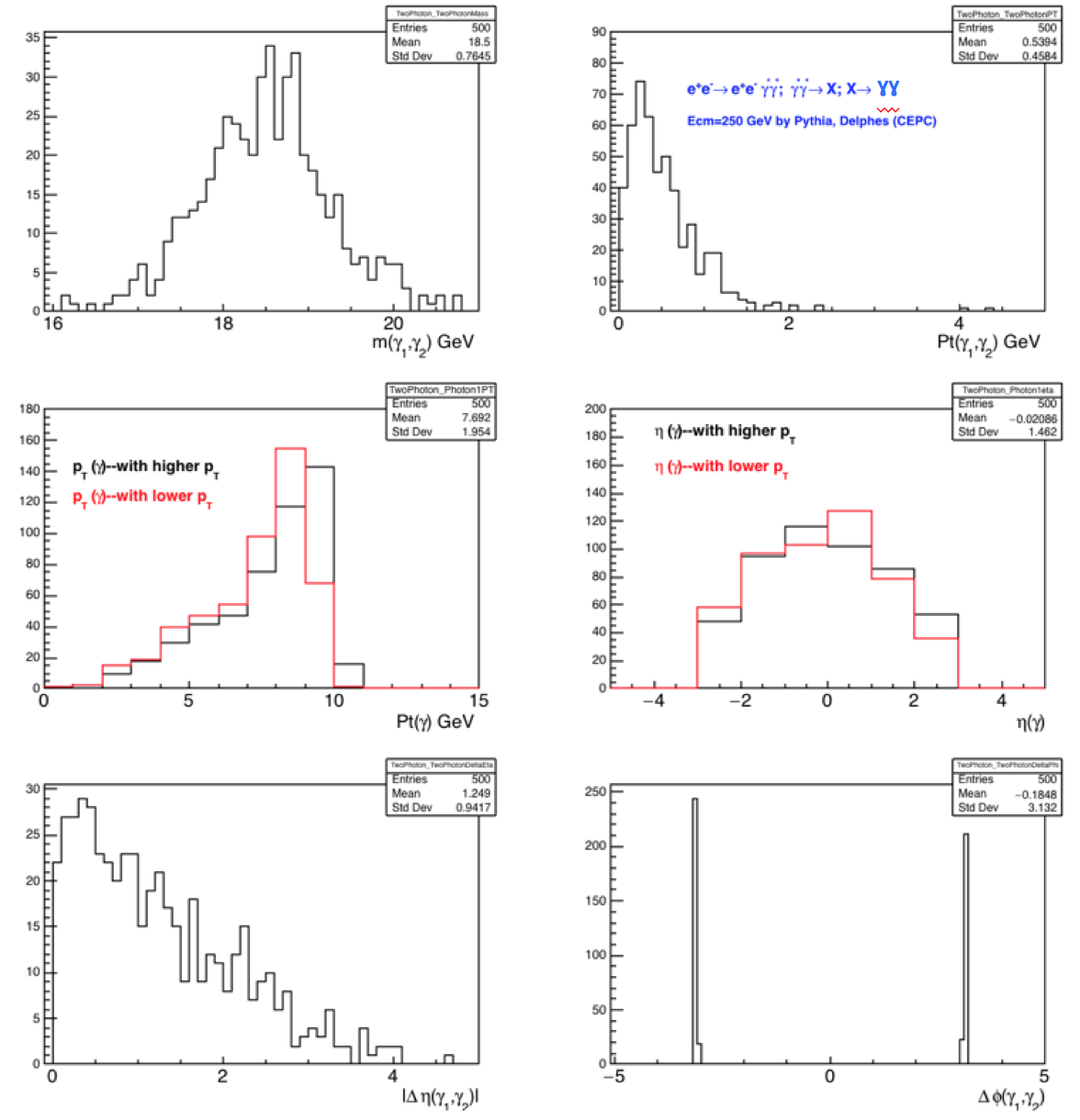}}
\caption{
The features of the $\gamma\gamma$ system in the $e^+e^- \rightarrow e^+e^- X; X\rightarrow \gamma\gamma$   
process 
at CEPC: 
(top left) The di-photon mass distribution;
(top right) The $p_T$ distribution of the di-photon;
(middle left) The $p_T$ distribution of each photon;
(middle right) The $\eta$ distribution of each photon;
(bottom left) The $\Delta \eta$ distribution of the di-photon;
(bottom right) The $\Delta \phi$ distribution of the di-photon.
These events are simulated by Delphes with CEPC configuration.
\label{cepcfigtvp3}}
\end{figure}

\begin{figure}[b]
\centerline{\includegraphics[width=15.0cm]{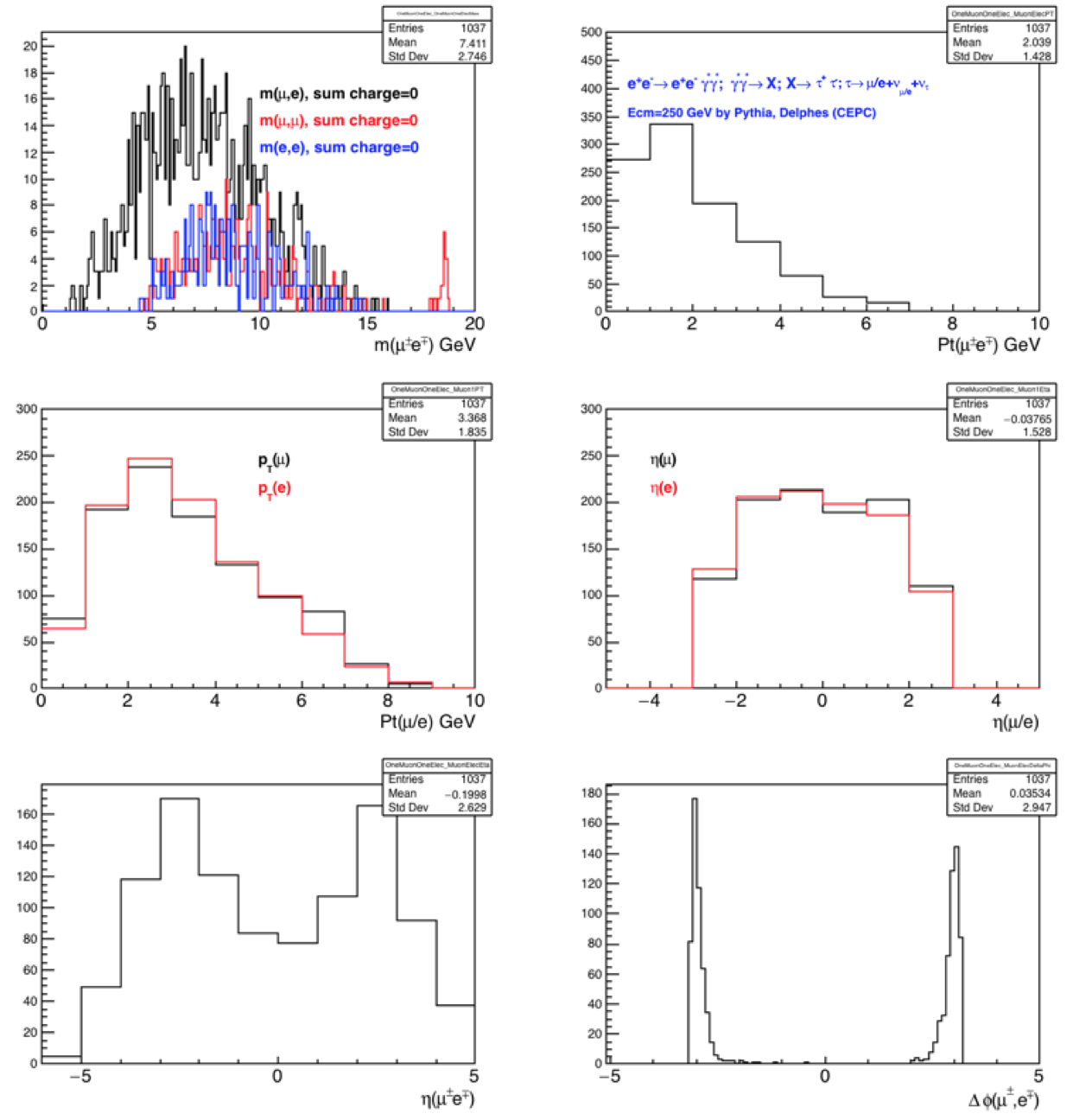}}
\caption{
The features of the $\mu^{\pm}e^{\mp}$ ($\tau^{\pm}\tau^{\mp}$) system 
in the $e^+e^- \rightarrow e^+e^- X; X\rightarrow \tau^{\pm}\tau^{\mp}; \tau^{\pm}\rightarrow\ell^{\pm}+anything-else, 
\tau^{\mp}\rightarrow \ell^{\mp}+anything-else$, where $\ell=\mu,e$ 
process 
at CEPC: 
(top left) The $\mu^{\pm}e^{\mp}$/$\mu^{\pm}\mu^{\mp}$/$e^{\pm}e^{\mp}$ mass distributions;
(top right) The $p_T$ distributions of $\mu^{\pm}e^{\mp}$;
(middle left) The $p_T$ distribution of muon and electron;
(middle right) The $\eta$ distribution of muon and electron;
(bottom left) The $\eta$ distribution of $\mu^{\pm}e^{\mp}$;
(bottom right) The $\Delta \phi$ distribution  between muon and electron.
These events are simulated by Delphes with CEPC configuration.
\label{cepcfigtvp4}}
\end{figure}

\begin{itemize}

\item $e^+e^- \rightarrow e^+e^- X; X\rightarrow Y(1S) \mu^+\mu^-; Y(1S)\rightarrow \mu^+\mu^-$.
Figure~\ref{cepcfigtvp1} shows the reconstructed mass, $\eta$, $p_T$, $p_z$ and $\phi$  
distributions for a potential 18 GeV object and its decay particles from the two-photon production. 
Since it is singly produced and there is no recoiling boost from associated 
particles in the transverse plane, the object has zero transverse 
momentum, and the transverse momentum of its decay particles are low.
The object is distributed in the forward region although its decay particles 
are in the central region along one side.
Figure~\ref{cepcfigtvp2}  shows the 
six $\Delta R$ distributions between the six pairs, they are well separated 
between each other, unlike the distributions from the associate production.

\item $e^+e^- \rightarrow e^+e^- X; X\rightarrow \gamma\gamma$.
Figure~\ref{cepcfigtvp3} shows the reconstructed mass, $p_T$, $\eta$,  and 
$\phi$  distributions for a potential 18 GeV object and its decaying particles. 
As a general feature from two-photon process, this object has a 
close to zero transverse momentum distribution, thus the photon transverse 
momentum distributions peak at high end around 9 GeV to make the object's mass.
The two decaying photons are back-to-back in $\phi$ plane, and 
close in $\eta$ plane along one side, thus distributed in the forward region.

\item $e^+e^- \rightarrow e^+e^- X; X\rightarrow \tau^{\pm}\tau^{\mp}; \tau^{\pm}\rightarrow\ell^{\pm}+anything-else, 
\tau^{\mp}\rightarrow \ell^{\mp}+anything-else$, where $\ell=\mu,e$.
Figure~\ref{cepcfigtvp4} shows the reconstructed $\mu^{\pm}e^{\mp}$, $\mu^{\pm}\mu^{\mp}$ 
and $e^{\pm}e^{\mp}$ masses, $p_T$, $\eta$, $\Delta \eta$ and 
$\Delta \phi$  distributions for the dilepton system. 
The  dilepton mass distributions are the same as that in the associate production:
widely peak below 18.5 GeV. 
The dilepton system  is distributed in the forward region but each lepton is  mostly 
distributed in the detector's central 
region along one side and back-to-back in $\phi$ spaces. 
Although it is diluted for the dilepton transverse momentum due to missing 
neutrino particles in the final state, the dilepton transverse momentum 
distribution is still concentrated at low momentum end (around 2 GeV), this can be 
used to separated signal events from background events. 

\end{itemize}

\clearpage
\newpage

\section{Outlook--a first low energy photon collider somewhere}

\begin{figure}[b]
\centerline{\includegraphics[width=15.0cm]{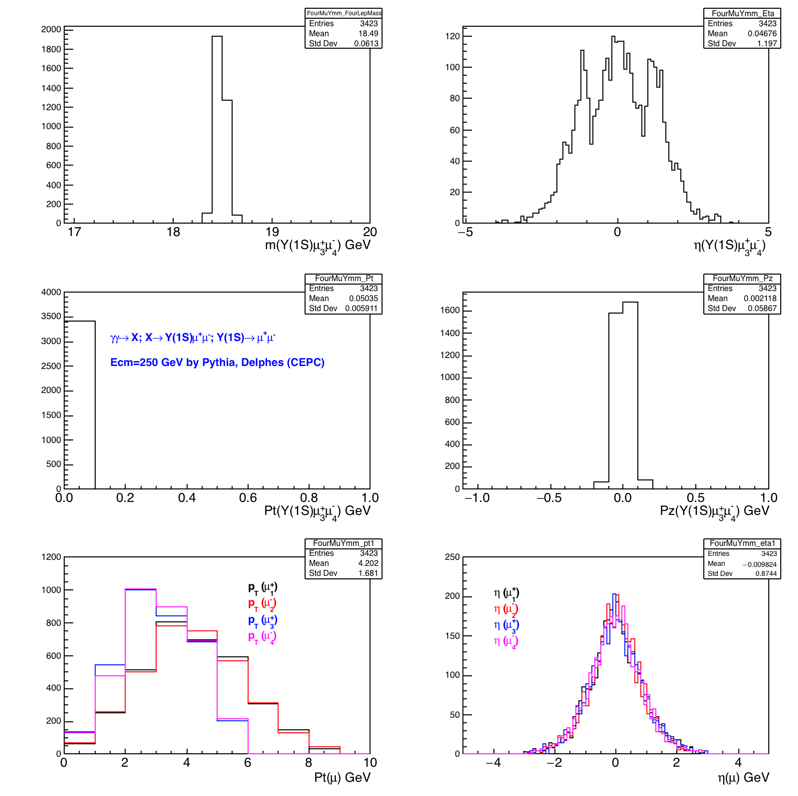}}
\caption{
The features of the four-muon system in a visionary photon collider with total energy of 18.5 GeV 
($\gamma\gamma \rightarrow X; X\rightarrow Y(1S) \mu^+\mu^-; Y(1S)\rightarrow \mu^+\mu^-$)  
(top left) The invariant mass; 
(top right) The $\eta$ distribution;
(middle left)$p_T$;
(middle right) $p_Z$;
(bottom left) $p_T$ of the four muons.
(bottom right) $\eta$ of the four muons.
These events are simulated by Delphes with CEPC configuration.
\label{photoncolliderfig1}}
\end{figure}

A detail overview about two-photon physics can be found at Ref.~\cite{ilya}, 
which summarizes the two-photon process at various colliders including 
possible photon colliders.  
There are many benefits to study states with spin 0 and spin 2 particles 
via a photon collider  such as high luminosity, studying possible CP 
violation through polarized photons etc.  
The kinematic distributions for various decays are almost  the same as that 
in the two-photon process at CEPC.  As an example, we use this process:
$\gamma \gamma \rightarrow X; X\rightarrow Y(1S) \mu^+\mu^-; Y(1S)\rightarrow \mu^+\mu^-$
to demonstrate its kinematic distributions as shown in 
Fig.~\ref{photoncolliderfig1}, where the total colliding energy is 18.5 GeV. 
The author is looking for possible collaboration to extend this section in the future.

\section*{Acknowledgments}

I would like to thank Stephen Mrenna, Stefan Prestel for their many detailed help 
on the usage of Pythia.
All opinions and comments expressed, and any error committed,
are solely responsibility of the author himself.



\end{document}